\documentclass[10pt,letterpaper]{article}
\usepackage{opex3}
\usepackage{amssymb}
\usepackage{cite}

\begin{document}
\newcommand{\ket}[1]
{\left|#1\right\rangle}

\newcommand{\bra}[1]
{\left\langle #1\right|}

\newcommand{\braket}[2]
{\left\langle #1\right|\left.#2\right\rangle}

\title{Displacement-enhanced entanglement distillation of single-mode-squeezed entangled states}

\author{Anders Tipsmark, Jonas S. Neergaard-Nielsen,$^{\ast}$ and Ulrik L. Andersen} 

\address{Department of Physics, Technical University of Denmark, Fysikvej, 2800 Kgs. Lyngby, Denmark}

\email{$^{\ast}$jsne@fysik.dtu.dk}


\begin{abstract}
It has been shown that entanglement distillation of Gaussian entangled states by means of local photon subtraction can be improved by local Gaussian transformations. Here we show that a similar effect can be expected for the distillation of an asymmetric Gaussian entangled state that is produced by a single squeezed beam. We show that for low initial entanglement, our largely simplified protocol generates more entanglement than previous proposed protocols. Furthermore, we show that the distillation scheme also works efficiently on decohered entangled states as well as with a practical photon subtraction setup.
\end{abstract}



\section*{}

Continuous variable (CV) entanglement is a valuable resource for many quantum informational protocols~\cite{Braunstein2005,Andersen2010,Weedbrook2012}. However, the performance of these protocols is often limited due to the difficulty in generating CV states with a high degree of entanglement. Moreover, even if a large degree of entanglement can be produced using a highly efficient nonlinear parametric process, the distribution of it (e.g. among two parties in a network) will inevitably lead to dissipation, rendering the state weakly entangled. To improve the performance of quantum information processing, it is therefore important to devise a protocol that increases the amount of entanglement between two distant parties by means of local quantum transformations and classical communication. This can be done by the process of entanglement distillation.

Distillation of non-Gaussian entanglement (either a pure or a mixed non-Gaussian entangled state) 
can be implemented using simple linear optics~\cite{Browne2003} as demonstrated in~\cite{Dong2008,Hage2008}. On the contrary, the distillation of Gaussian states (either pure or mixed Gaussian states) is challenging as it will inevitably require some non-Gaussian transformations~\cite{Eisert2002,Giedke2002,Fiurasek2002} enabled by a very strong Kerr nonlinearity~\cite{Duan2000a,Fiurasek2003}, using a non-Gaussian measurement~\cite{Opatrny2000,Cochrane2002,Olivares2003}, using non-Gaussian resources~\cite{Xiang2010} or using a combination of photon addition and subtraction~\cite{Yang2009,Fiurasek2010,Lee2011c,Navarrete-Benlloch2012}. An intriguing scheme for entanglement distillation of Gaussian states was suggested by Opatrn\'y et al. \cite{Opatrny2000} and involves local photon subtraction of a two-mode entangled state. The scheme was recently implemented by Takahashi et al~\cite{Takahashi2010a} (entanglement distillation by non-local photon subtraction has also been demonstrated~\cite{Ourjoumtsev2007b}). This photon subtraction scheme can however be improved by local Gaussian operations: Zhang and van Loock~\cite{Zhang2011} showed that local squeezing operations prior to photon subtraction improve the performance of distillation in terms of producing states with a higher degree of entanglement and with higher success rate. Secondly, it was realized by Fiur\'a\v{s}ek~\cite{Fiurasek2011a} that by using the much simpler operations of local phase space displacements, a similar improvement can be achieved.


In all these previous proposals on entanglement distillation, the considered Gaussian entangled state was produced by interfering two single mode squeezed states on a beam splitter. However, in practice it is much simpler to generate entanglement from a single squeezed mode that is split on a balanced beam splitter, as was done in \cite{Takahashi2010a}. It was found in \cite{Tipsmark2012,Cernotik2012} that -- surprisingly -- in some cases the usage of a single mode squeezed beam for the generation of Gaussian entangled states, the degree of entanglement after distillation is higher than if a two-mode squeezed state was used. In this paper, we further analyze the displacement enhanced entanglement distillation with a focus on the initial single mode squeezed state, which will be closer to an experimental realization. We investigate the photon number distribution of the distilled states to gain further insight into how displacement helps, and we study the influence of losses in the distribution channels on the attainable entanglement and optimal displacements.

\begin{figure}[t]
\begin{center}
\includegraphics[width=\textwidth]{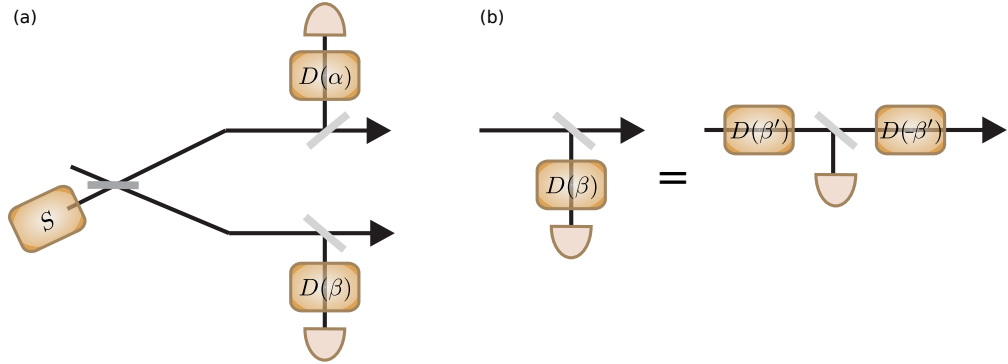}
\caption{a) Schematic of the proposed setup. $S$ denotes the single mode squeezed state and $D(\alpha),D(\beta)$ are displacement operations. b) Two equivalent setups for displacement controlled photon subtraction where $\beta' = \sqrt{1-T} \beta$ and $T$ is the transmission of the beam splitter.}
\label{fig:setup}
\end{center}
\end{figure}

We consider the entanglement distillation setup shown in Fig.~\ref{fig:setup}. The entangled state is simply produced by dividing a single mode squeezed state on a balanced beam splitter, and the resulting modes are sent through lossy channels to the two sites, denoted A and B. At these two sites a Gaussian displacement transformation as well as single photon subtractions are applied to distill the quantum state. In the following we first consider the distillation protocol when the channels are loss-free and the photon subtraction is ideal, and secondly we consider the more realistic scenario where the channels are lossy and the photon subtraction process is non-ideal.  

%
%
%
%

\begin{figure}[t]
\begin{center}
\includegraphics[width=\textwidth]{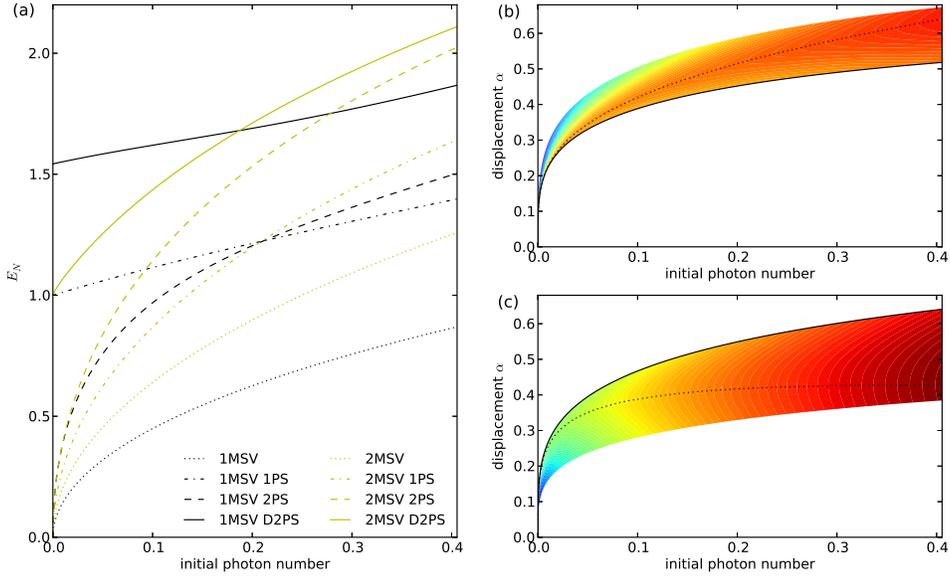}
\caption{(a) Logarithmic negativity as a function of the average number of photons of the initial squeezed state. 1(2)MSV: Single (two) mode squeezed vacuum states used initially. 1(2)PS: Single (two) photon subtraction. D2PS: Displacement based two-photon subtraction with optimal displacement. 
(b) The logarithmic negativities obtained in the 1MSV D2PS setting for varying displacement amplitudes $\alpha=-\beta$. The color scale goes from $E_N=0$ (blue) to $E_N=2.1$. The dashed black curves indicate the  $\alpha$ values that optimize $E_N$ (which results in the entanglement curve in (a)), while the solid line follows the optimal value in the low-squeezing limit of Eq. (\ref{disp}).
(c) As (b), but for the 2-mode squeezed vacuum.
}
\label{fig:negativity}
\end{center}
\end{figure}

Our starting point is a single mode squeezed state mixed with vacuum on a balanced beam splitter (written in the Fock state representation): 
\begin{eqnarray}
\ket{\zeta_1}~&=&~U_{BS}(1-\gamma_1^2)^{1/4}\sum_{n=0}^\infty\frac{\sqrt{(2n)!}}{n!}\left(\frac{\gamma_1}{2}\right)^n\ket{2n}|0\rangle,	\label{SMS}
\end{eqnarray}
where $U_{BS}$ is the beam splitter unitary transformation, $\gamma_1=\tanh(s_1)$ and $s_1$ is the squeezing parameter.  
In the weak squeezing limit ($\gamma\ll 1$) the state can be truncated to 
\begin{eqnarray}
\ket{\zeta_1}~&\approx&~\ket{00}+\frac{\gamma_1}{\sqrt{2}}\left(\frac{1}{2}\ket{20}+\frac{1}{\sqrt{2}}\ket{11}+\frac{1}{2}\ket{02}\right).	\label{SMS2}
\end{eqnarray}
The entanglement of a pure state like this is usually quantified by the entropy of entanglement; however, since we will later compare with mixed states for which the entropy is not a proper measure, we adopt instead as our entanglement measure the logarithmic negativity (LN). The LN is defined as the binary logarithm of the trace norm of the partially transposed density matrix $(\ket{\zeta_1}\bra{\zeta_1})^{T_A}$ \cite{Vidal2002}. The logarithmic negativity of the above state as a function of the average photon number $\langle n \rangle = \gamma_1^2/(1-\gamma_1^2)$ is plotted in Fig.~\ref{fig:negativity}(a) by the black dotted line. 

The idea is now to increase the entanglement of the state, and this can be done by removing the vacuum contribution and equalizing the weight of the low order excitation terms. For weak squeezing (as in Eq.~(\ref{SMS2})), it is straightforwardly realized that such an equalization can be obtained simply by performing single photon subtraction on just one of the modes A or B. For single photon subtraction of mode A, enabled by the bosonic annihilation operator $a_A$, the result is
\begin{eqnarray}
\ket{\zeta_1}~&\rightarrow&~\hat{a}_A\otimes\mathbb{I}_B\ket{\zeta_1}\\
~&=&~\frac{\gamma_1}{2}\left(\ket{10}+\ket{01}\right),	\label{SPS}
\end{eqnarray}
which is maximally entangled (in the two-dimensional subspace, thus neglecting higher order terms). The resulting entanglement is plotted in Fig.~\ref{fig:negativity}(a) with the black dash-dotted line, and it is shown that for very low squeezing degrees the LN reaches the maximum value of 1 for a 2D Hilbert space. However, the state should be described in a larger Hilbert space in which the state is not maximally entangled.
We next consider the simultaneous subtraction of single photons at both sites as described by the operator $\hat{a}_A\otimes\hat{a}_B$ and denoted 2PS (2-Photon-Subtraction). The resulting LN is plotted in Fig. \ref{fig:negativity}(a) by the black dashed line. In contrast to the previous protocol (with a single photon being subtracted, 1PS), this protocol is not very effective for low average photon numbers but for higher numbers (larger than about 0.21, corresponding to 3.9 dB of squeezing) it becomes more effective. 

The entanglement can however be further enhanced for all degrees of initial squeezing by applying a Gaussian transformation prior to photon subtraction. The transformation that leads to this enlargement is the simple phase space displacements, $D(\alpha)=\exp(\alpha a_A^\dagger-\alpha^* a_A)$ and $D(\beta)=\exp(\beta a_B^\dagger-\beta^* a_B)$, where $\alpha$ and $\beta$ are the complex excitations of the displacements. By implementing these displacements in mode A and B prior to and after the two photon subtractions (see Fig.~\ref{fig:setup}(b)), the state reads 
\begin{eqnarray}
\ket{\Psi}~&=&~\left(\alpha\beta+\frac{\gamma_1}{2}\right)\ket{00}+\frac{\gamma_1}{\sqrt{2}}(\beta+\alpha)\frac{\ket{10}+\ket{01}}{\sqrt{2}}\nonumber\\	&&+\frac{\gamma_1}{\sqrt{2}}\alpha\beta\left(\frac{1}{2}\ket{20}+\frac{1}{\sqrt{2}}\ket{11}+\frac{1}{2}\ket{02}\right)+O(\gamma_1^2).\nonumber
\end{eqnarray}
It is clear from this expression that the vacuum and single-photon contributions, which prevent the state from being strongly entangled, can be removed by setting 
\begin{eqnarray}
	\alpha~=~-\beta~=~\sqrt{\frac{\gamma_1}{2}},
\label{disp}
\end{eqnarray}
which can be shown to give 
\begin{eqnarray}
\ket{\Psi}~=~-\frac{\gamma_1^2}{2\sqrt{2}}\left(\frac{1}{2}\ket{20}+\frac{1}{\sqrt{2}}\ket{11}+\frac{1}{2}\ket{02}\right)+O(\gamma_1^3).	\label{2PSD}
\end{eqnarray}
This state contains an entanglement of $E_N=1.54$ for low initial squeezing degrees. This is not maximally entangled in the 3-dimensional Hilbert space due to the unequal weights but it is close to - the maximally entangled state would have $E_N=1.585$.  

As can be seen from Fig.~\ref{fig:negativity}(b), the displacements in Eq.~(\ref{disp}) are maximizing the LN only for low degrees of initial squeezing. For larger squeezing degrees, the optimum displacement is larger than the one in Eq.~(\ref{disp}), and thus it is not optimized by the removal of the vacuum term. The single-photon components are however always cancelled by the choice of $\alpha=-\beta$. 
The resulting optimized LN is plotted in Fig.~\ref{fig:negativity}(a) (both by black solid curves).
Clearly, the pre-Gaussian processing improves the entanglement for all average photon numbers~\cite{Tipsmark2012}. From Fig.~\ref{fig:negativity}(b) we furthermore see that the entanglement is highly sensitive to the exact value of the displacement amplitude for small initial photon numbers, but less so for increasing photon numbers.

%
%
%
%

For comparison, we also briefly consider the distillation of a two-mode squeezed state as was treated in \cite{Fiurasek2011a}. Here the starting point is 
\begin{eqnarray}
	\ket{\zeta_2}_{AB}~&=&~\sqrt{1-\gamma_2^2}\sum_{n=0}^\infty\gamma_2^n\ket{n}_A\ket{n}_B	
	\label{TMS}
\end{eqnarray}
where $\gamma_2=\tanh(s_2)$ and $s_2$ is the two-mode squeezing parameter. This state, which has an average photon number $\langle n \rangle = 2\gamma_2^2/(1-\gamma_2^2)$, can be produced by interfering and phase locking two single mode squeezed vacua on a balanced beam splitter. The implementation of a local single photon subtraction, either on one or both sites, transforms the entangled state into another entangled state with higher LN as illustrated in Fig. \ref{fig:negativity}(a) by yellow curves. Note that for identical initial $\langle n \rangle$, $\gamma_2$ is smaller than $\gamma_1$.
By displacing the state with excitations of $\alpha_A~=~-~\alpha_B~=~\sqrt{\gamma_2}$ before subtracting two photons, the distilled state reads $\ket{\psi^{\gamma_2}}~=~\gamma_2^{3/2}(\ket{10}-\ket{01})+O(\gamma_2^2)$ for weak initial entanglement ($\gamma_2 \ll 1$). This state is also maximally entangled in the two-dimensional sub space for low initial squeezing as was the case for the 1PS single mode squeezed state in Eq. (\ref{SPS}). 
For larger initial squeezing levels, the optimal displacement is lower than $\sqrt{\gamma_2}$, as seen from Fig. \ref{fig:negativity}(c). The sensitivity to the displacement amplitude is however less than for the single-mode squeezed input state.

In particular, we note that the degree of entanglement after distilling the two-mode state using the displacement-enhanced protocol is lower than the one obtained for the single mode scheme if the average photon number is low. That is, using the largely simplified protocol with a single mode squeezer split on a beam splitter (rather than interfering and phase locking two single-mode squeezers) the distillation transformation produces an entangled state that contains more entanglement. However, this conclusion only holds for a low average photon number. For higher photon numbers, the usage of a two-mode squeezed state results in a state with larger entanglement than if a single-mode squeezed state was used. But the low-gain regime is often of interest, since entanglement in this region is fairly easy to prepare, manipulate and maintain. 
 
%
%
\begin{figure}[t]
\begin{center}
\includegraphics[width=\textwidth]{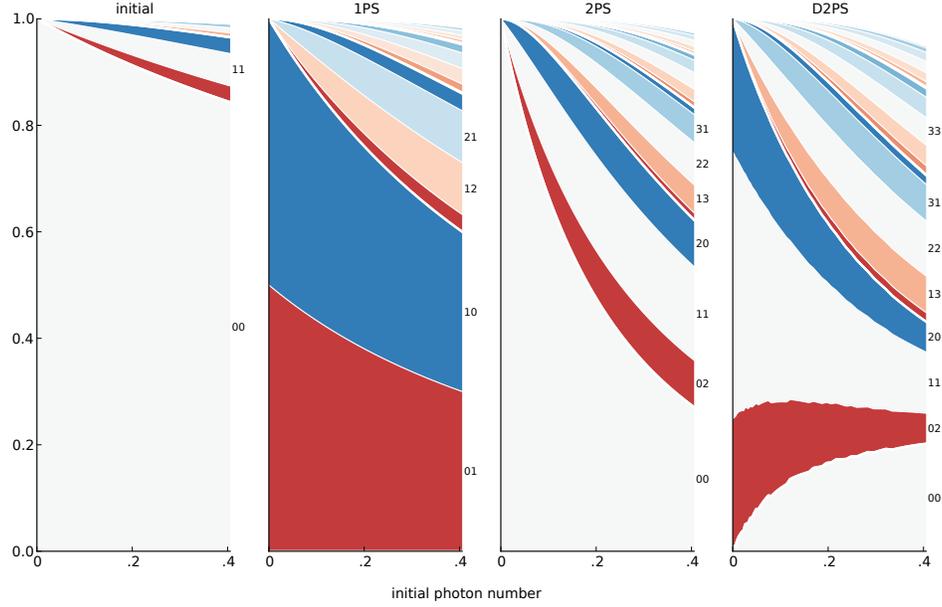}
\caption{Distribution of the different eigenstates of the entangled state that is produced from a single squeezed mode. The distributions are for the initial state (before distillation), the single photon subtracted state (1PS), the two photon subtracted state (2PS) and the displacement-enhanced protocol.}
\label{distri}
\end{center}
\end{figure} 
%
%

It is known that entanglement is optimized when the dominant eigenstates of a given state have equal weights as is the case in Eq.~(\ref{SPS}) and partially in Eq.~(\ref{2PSD}). To get some physical insight into the formation of stronger entanglement, in Fig.~\ref{distri} we plot the weights of the different eigenstates resulting from the different distillation protocols using a single mode squeezed state. (These, as well as most of our other numerical calculations were done by the package \cite{Johansson2012}.) Due to the very large vacuum contribution in the initial entangled state, it is evident that this state is not highly entangled. However, by implementing the 1PS scheme, the vacuum term is basically split in two new eigenstates ($|01\rangle$ and $|10\rangle$) leading to a large increase in the entanglement. By subtracting two photons (2PS scheme) rather than a single photon, the balancing of the eigenstates is destroyed for low photon numbers as the vacuum term again becomes dominant, and thus the entanglement is lower. However, by implementing the displacement-enhanced scheme (D2PS), the weights between the eigenstates are partially re-balanced, thereby producing strong entanglement. 

Based on this discussion, we can also clarify the reason behind a somewhat unintuitive observation from Fig. \ref{fig:negativity}(a), namely that a large amount of entanglement in some cases can be extracted from even very weakly squeezed initial states. As we just saw, the subtraction of a single photon or two photons following a displacement actually \textsl{adds} non-local photons to the final, heralded state. This is of course only possible because the photon subtraction is a highly probabilistic process -- see the discussion on success probabilities later. With the two-mode squeezed vacuum as initial state, a delocalized photon is also obtained in the weak squeezing limit with displacement and two-photon subtraction, giving a large entanglement. For the single-photon subtraction, however, no entanglement is obtained -- in stark contrast to the 1MSV case. The origin of this difference is the different form of the initial states: For 1MSV, the state in Eq.~(\ref{SMS2}) is a delocalized 2-photon state (plus vacuum), which results in a delocalized single photon after subtraction. For 2MSV, Eq.~(\ref{TMS}) is a twin-photon state (plus vacuum), which turns into a \textsl{localized} single photon upon detection in one of the modes.

%
%
\begin{figure}[b]
\begin{center}
\includegraphics[width=\textwidth]{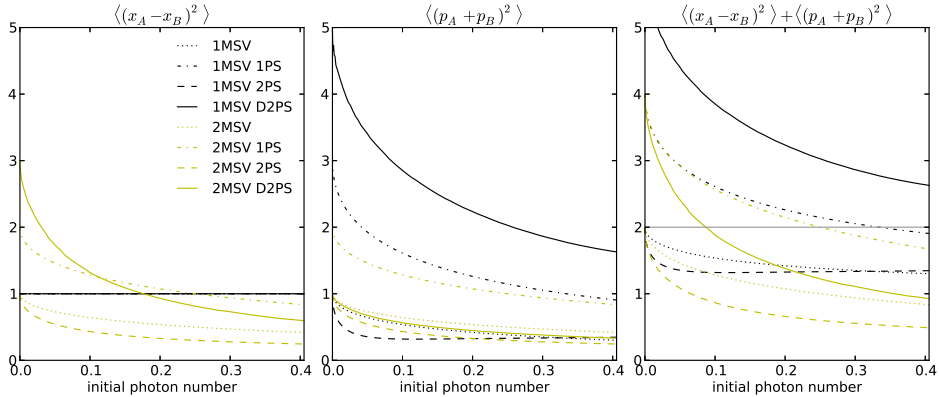}
\caption{Two-mode squeezing: Variance of the $x$-quadrature difference (left), 
         $p$-quadrature sum (center) and their sum (right) for the same states
         as studied in Fig. \ref{fig:negativity}(a). See that figure for abbreviations.
         The gray line at 2 in the
         right graph indicates the Simon-Duan entanglement criterion for Gaussian
         states. In these plots, however, only the 1MSV and 2MSV states are Gaussian.}
\label{fig:variance}
\end{center}
\end{figure} 
%
%

The photon subtraction is a highly non-Gaussian process, so although the initial entangled state is Gaussian this is not the case for the distilled state. A commonly used necessary and sufficient criterion for entanglement of two-mode Gaussian states (after suitable local transformations) is that of Simon \cite{Simon2000} and Duan et al \cite{Duan2000}: The state is entangled when the sum of the variances of the $x$-quadrature difference and $p$-quadrature sum of the two modes (the two-mode squeezing) is below the corresponding vacuum noise level, $\langle \Delta x_-^2 \rangle + \langle \Delta p_+^2 \rangle < 2$. Here, $x=(a^\dag+a)/\sqrt2$, $p=i(a^\dag-a)/\sqrt2$ and $x_-=x_A-x_B$, $p_+=p_A+p_B$. This criterion does not apply to non-Gaussian states, but it may still be relevant to see how the distilled states evaluate on this variance metric. For continuous variable quantum teleportation, for example, the performance is determined by the amount two-mode squeezing. We therefore plot the two-mode squeezing and its $x$ and $p$ components in Fig. \ref{fig:variance}.
One can see the simpler structure of the single-mode squeezed state in that $\langle \Delta x_-^2 \rangle$ is at the vacuum level for all cases, with and without photon subtraction. It is noteworthy from the right hand graph that, while the initial Gaussian states and the states distilled by non-displaced two-photon subtraction are always squeezed below the vacuum level, the situation is much more complex for the cases of one-photon subtraction and two-photon subtraction with displacement. When just considering the second-order moments as we do here, these states appear very ``noisy'' for low initial photon numbers, while they become squeezed for higher photon numbers. This is in spite of their large entanglement as quantified by the logarithmic negativity in Fig. \ref{fig:negativity}. Most remarkably, the 1MSV D2PS state which contains the most entanglement for low initial photon numbers is also by far the state with the largest variance due to its non-Gaussian character. This means that, although it is highly entangled, it will not be the optimal choice for all protocols. It may, however, be possible to turn the state more Gaussian through a protocol like the one in \cite{Browne2003}.

\begin{figure}[t]
\begin{center}
\includegraphics[width=\textwidth]{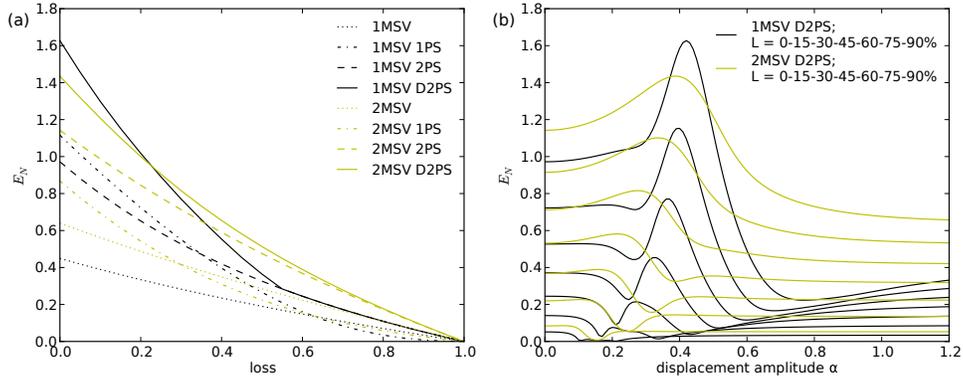}
\caption{(a) The logarithmic negativity is plotted as a function of the loss of the two channels. The attenuation is identical in the two channels. See caption of Fig.~\ref{fig:negativity} for abbreviations. 
(b) The logarithmic negativity as a function of the displacement amplitude for the two-photon subtracted single- and two-mode squeezed vacuum with losses varying from 0 (top curves) to 0.9 (bottom curves) in steps of 0.15. In both panels, the initial photon number is 0.1.}
\label{fig:loss}
\end{center}
\end{figure}

We now investigate the performance of the protocol under dissipation. This is of interest as in most protocols the entangled state is distributed in a network connected by lossy channels, thereby rendering the entangled states in any practical realization impure. The logarithmic negativities after distillation using the different strategies outlined above for the single- and two-mode squeezed states are depicted in Fig.~\ref{fig:loss}(a) for $\langle n\rangle=0.1$ as a function of the channel attenuation. It is assumed that the two channels possess identical attenuations. 
Of course, the LN decrease with losses, but it appears that the displacement-distilled single-mode squeezed vacuum is more fragile to attenuation than the two-mode squeezing. While it is the most strongly entangled state in the ideal case, the exposure to losses soon makes it less entangled than the two-photon subtracted two-mode squeezing, both with and without displacement. Moreover, as the losses increase further, a point is reached where it is no longer advantageous to do displacement before the photon subtraction -- the kink on the curve around 55\% loss. This behaviour is not observed for the 2MSV input.
  
%
%
%
%
%
%
\begin{figure}[t]
\begin{center}
\includegraphics[width=\textwidth]{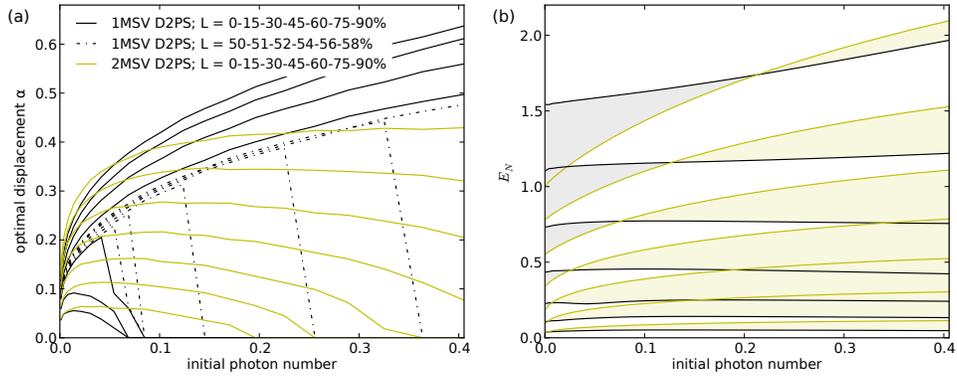}
\caption{(a) Optimal displacement as a function of channel losses for varying losses. The losses are increasing for lower-lying curves. (b) Maximum attainable logarithmic negativity for varying losses. The losses are the same as for the solid curves in (a). The shaded areas in gray (yellow) designate the ranges of initial photon numbers for which the 1MSV (2MSV) states are superior.}
\label{fig:loss_squeezing}
\end{center}
\end{figure} 
%
%

To investigate it further, we calculate in Fig.~\ref{fig:loss}(b) the LN for the displacement distillation as a function of $\alpha$ for a range of different losses. The trends are qualitatively different for the 1MSV and 2MSV states. Whereas the 2MSV states have distinct optima that gradually tend towards zero as losses increase, the location of the peaks of the 1MSV curves go more slowly towards smaller amplitudes. On the other hand, for high losses the height of the peaks dip below the values for zero displacement -- this is where the kink in the 1MSV D2PS curve in (a) comes from. Consequently, we discover that this curve in fact consists of two separate regimes: To the left of the kink on the low-loss side, the entangled states have been distilled \textsl{with} displacement, while on the high-loss side, the distillation took place \textsl{without} displacement. We would therefore expect the states to be of very different character, for example in terms of their quadrature variance -- referring to Fig. \ref{fig:variance}(c), the two different regimes are, respectively, above and below the vacuum level -- and this is indeed the case (not displayed here).

In Fig.~\ref{fig:loss_squeezing}(a) we take a complementary look at the optimal distillation strategy. The optimal displacement amplitudes (those maximizing the LN) are plotted as a function of initial photon number for the same range of losses as in Fig.~\ref{fig:loss}(b). For zero loss, the 1MSV and 2MSV curves are identical to those in Figs.~\ref{fig:negativity}(b,c). We see the same pattern as observed before: For 2MSV initial state, the optimal displacement decreases gradually to zero with incresing losses. Interestingly, though, for large initial squeezing levels the displacement ceases to be effective at around 50-80\% channel attenuation, while for lower squeezing levels it is always beneficial to implement a displacement prior to photon detection. 
While this latter observation is also true for the 1MSV case, the transition from the displacement-enhanced regime to the regime where it is optimal to avoid the displacement (the ``kink'') happens drastically, rather than smoothly as for 2MSV.  The transition takes place for channel losses just above 50\% as seen from the extra dashed curves.

If we disregard the added complexity of using an initial 2MSV state instead of 1MSV, it is of interest to consider which of the two initial state preparations provide the most entanglement after the distillation process. As we have seen, for small to moderate photon numbers the 1MSV state can be distilled to the highest level of logarithmic negativity. If the distribution channels are lossy, however, we saw from Fig.~\ref{fig:loss}(a) that the 1MSV degrades faster. In Fig.~\ref{fig:loss_squeezing} we plot the attainable LN for our representative range of losses. It is evident that with increasing attenuation, using 2MSV rapidly becomes advantageous for almost all squeezing levels. Moreover, it also appears that for almost any amount of loss, there is no benefit at all in terms of LN of increasing the squeezing/photon number of the 1MSV state -- the levels are essentially constant. Of course, the event rate will still be higher with higher initial photon numbers.

In the above study we have assumed that a single photon is perfectly removed from each of the entangled modes, and we have modelled this by the annihilation operator. This is however an idealization. In practice a photon is usually subtracted by reflecting a small part of the beam on an asymmetric beam splitter, and subsequently measuring the presence of a single photon in the reflected mode. The most practical detector for this purpose is an avalanche photodiode (APD) which is an on/off detector that discriminates between zero photons and some photons. Assuming that the reflectivity of the asymmetric beam splitter as well as the initial average photon number are low, the reflected beam will contain much less than a single photon on average, and thus the APD will effectively work as a single photon counter. On the other hand, for a finite reflectivity and a large initial squeezing, in some rare (although non-negligible) cases two photons will impinge onto the detector which cannot be discriminated from a single photon event. This causes an error. 

%
%
\begin{figure}[t]
\begin{center}
\includegraphics[width=\textwidth]{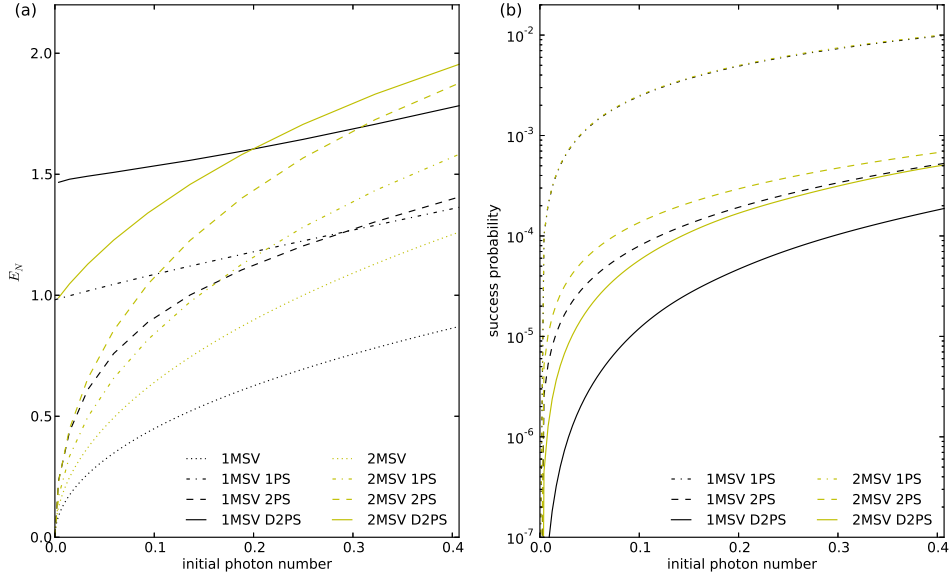}
\caption{Performance of the distillation protocol for realistic photon subtraction. Logarithmic negativity (a) and success probability (b) as a function of the initial average photon number. The tap-off beamsplitter reflectivity for photon subtraction is 5\%. See caption of Fig.~\ref{fig:negativity} for abbreviations.}
\label{neg_onoff}
\end{center}
\end{figure} 
%
%

To simulate the realistic setup, we have used a more rigourous model in which the initial state (either Eq.~(\ref{SMS}) or Eq.~(\ref{TMS})) is transformed through a beam splitter with a reflectivity of 5\%, displaced and finally measured with the projector $\Pi=1-|0\rangle\langle 0|$.  The results (both for the single mode and the two-mode squeezed state) are shown in Fig.~\ref{neg_onoff}. It is obvious from the figures that the trend of the LN as a function of the initial average photon numbers is identical to the trend for the ideal distillation scheme in Fig.~\ref{fig:negativity}(a), but the amount of entanglement is slightly lower in the former case. 
Another crucial parameter for characterizing the performance of the distillation protocols is the success probability. This is plotted in Fig.~\ref{neg_onoff}(b). At first it seems counter-intuitive that the success probability decreases when displacement is included, but it can be understood as follows. 
The rate of the initial photon detection in mode A is indeed increased when displacement is introduced. However, photon subtraction from a squeezed state \textsl{increases} its average photon number, so the displacement (which is experimentally implemented by admixture of a coherent state) leads to a \textsl{lower} increase in the photon number of mode B after the subtraction in mode A. Furthermore, the displaced photon subtraction results in a state which has a small displacement in phase space (as opposed to the zero-mean of the initial state). The subsequent displacement of mode B before the photon detection there is opposite in direction of the state's displacement, leading to destructive interference in the detected mode. As a result of these two effects, the success probability of the second photon detection is considerably lower with displacement than without, outweighing the increased probability of the first detection.

In conclusion, we have theoretically investigated a displacement-enhanced distillation scheme of entangled states that are produced by a single squeezed mode. We have found that a simple Gaussian displacement operation prior to photon subtraction increases the entanglement of the distilled state. Similar conclusion has been found for the two-mode squeezed state scheme, but in contrast to the previous proposals, the experimental realization of our scheme is much simpler as it does not require the control and phase locking of two independent squeezed beams. An experimental realization is therefore feasible with current technology~\cite{Tipsmark2011,Neergaard-Nielsen2010,Lee2011}. 
On the other hand, our analysis also shows that if the entanglement distribution channels are sufficiently lossy, it is still advantageous to use two-mode squeezing at the initial stage. This may also be required if a Gaussian-like two-mode squeezing is required for a given protocol.

\section*{Acknowledgments}
We acknowledge financial support from the Danish Agency for Science Technology and Innovation.

\end{document}